\documentclass[journal,comsoc]{IEEEtran}
\usepackage[T1]{fontenc}

\usepackage{tabularx}
\usepackage{xcolor}
\usepackage{multirow}
\usepackage{booktabs}
\usepackage{array}
\newcolumntype{L}[1]{>{\raggedright\let\newline\\\arraybackslash\hspace{0pt}}m{#1}}
\newcolumntype{C}[1]{>{\centering\let\newline\\\arraybackslash\hspace{0pt}}m{#1}}
\newcolumntype{R}[1]{>{\raggedleft\let\newline\\\arraybackslash\hspace{0pt}}m{#1}}

\usepackage{cite}

\ifCLASSINFOpdf
   \usepackage[pdftex]{graphicx}
\else
   or other class option (dvipsone, dvipdf, if not using dvips). graphicx
   \usepackage[dvips]{graphicx}
   \graphicspath{{../eps/}}
   \DeclareGraphicsExtensions{.eps}
\fi
\usepackage{amsmath}
\usepackage[cmintegrals]{newtxmath}
\usepackage{algorithmic}
\usepackage{algorithm}

\usepackage{array}
\usepackage{makecell}

\ifCLASSOPTIONcompsoc
  \usepackage[caption=false,font=normalsize,labelfont=sf,textfont=sf]{subfig}
\else
  \usepackage[caption=false,font=footnotesize]{subfig}
\fi
\begin{document}

\title{Sub-Nyquist Sampling OFDM Radar}
%
%

\author{Kawon~Han,~\IEEEmembership{Member,~IEEE,}
        Seonghyeon~Kang,~\IEEEmembership{Student Member,~IEEE,}
        and~Songcheol~Hong,~\IEEEmembership{Member,~IEEE}

\thanks{Manuscript received xx, 2023. This work was supported by the Institute of Information \& Communications Technology Planning and Evaluation (IITP) funded by the Korean Government, Ministry of Science and ICT (MSIT) under Grant 2019-0-00826.

Kawon Han is with the Interuniversity Microelectronics Centre (IMEC), 3001 Leuven, Belgium (e-mail: hkw1176@gmail.com).

Seonghyeon Kang and Songcheol Hong are with the Department of Electrical Engineering, Korea Advanced Institute of Science and Technology
(KAIST), Daejeon 34141, South Korea (e-mail: stx1226@kaist.ac.kr;songcheol1234@gmail.com).}}

\maketitle

\begin{abstract}
In this paper, we propose a sub-Nyquist sampling (SNS) orthogonal frequency-division multiplexing (OFDM) radar system capable of reducing the analog-to-digital converter (ADC) sampling rate in OFDM radar without any additional manipulations of its hardware and waveform. To this end, the proposed system utilizes the ADC sampling rate of ${B/L}$ to sample the received baseband signal with a bandwidth of $B$, where ${L}$ is a positive proper divisor of the number of subcarriers. This divides the baseband signal into $L$ sub-bands, folding into a sub-Nyquist frequency band due to aliasing. By leveraging known modulation symbols of the transmitted signal, the folded signal can be unfolded to the full-band signal. This allows an estimation of target ranges with the range resolution of the full signal bandwidth $B$ without the degradation of the maximum unambiguous range. During the signal-unfolding process, the signals from other sub-bands remain as symbol-mismatch noise (SMN), which significantly degrades the signal-to-noise ratio (SNR) of the detected targets. It also causes weaker targets to be submerged under the noise in range profiles. To resolve this, a symbol-mismatch noise cancellation (SMNC) technique is also proposed, which reconstructs the interfering signals from the other sub-bands using the detected targets and subtracts them from the unfolded signal. As a result, the proposed sub-Nyquist sampling OFDM radar and corresponding signal processing technique enable a reduction in the ADC sampling rate by the ratio of $L$ while incurring only a $10\log_{10}{L}$ increase in the noise due to noise folding. This is validated through simulations and measurements with various sub-sampling ratios.
\end{abstract}


%
\IEEEpeerreviewmaketitle

\section{Introduction}
%
%
%
%
\IEEEPARstart{T}HE exploitation of a digitally modulated waveform adds promising features to a millimeter-wave (mmWave) radar sensor. Compared to systems that use analog waveforms such as the frequency-modulated continuous wave (FMCW), digital radar using a phase-modulated continuous wave (PMCW) or orthogonal frequency division multiplexing (OFDM) has great potential for use in future radar systems owing to its  diversity of waveforms \cite{Digi1, Digi2, Digi3}. For multiple-input multiple-output (MIMO) radar, the waveform diversity of digital radar provides design flexibility for multiple transmit signals \cite{Digi4}. This also allows the use of multiple sensors in a networked radar system by mitigating mutual interference between the sensors \cite{Digi5}. Moreover, recent advances in integrated sensing and communication (ISAC) systems bring the convergence of RF systems, with mmWave digital radar being much more compatible than its analog counterpart \cite{ISAC1, ISAC2, ISAC3, ISAC4, ISAC5}.

Among digitally modulated waveforms, OFDM, mainly adopted in wireless communications, offers all of the aforementioned advantages associated with a digital radar system \cite{OFDM1, OFDM2, OFDM3}. Nevertheless, the OFDM radar has several limitations that must be overcome before it can be implemented with high-resolution and low-cost sensors. One of the main constraints is that a large baseband bandwidth, equal to the RF bandwidth, is required \cite{OFDM4, OFDM5, OFDM6}. This implies that the sampling rate of the analog-to-digital converter (ADC) must be large enough to reconstruct the full-band signal. To provide a range resolution that exceeds 10 cm, the signal bandwidth as well as the ADC sampling rate should exceed 1.5 GHz. Moreover, the utilization of MIMO radar increases the number of baseband channels further, which greatly increases the ADC sampling burden. Currently, it is ineffective in terms of both cost and power to implement a fully digital high-resolution radar system \cite{Angle1, Angle2}.    

To address this issue, several systematic efforts have been made to reduce the baseband sampling rate in OFDM radar \cite{SC_OFDM1, SC_OFDM2, SC_OFDM3, FC_OFDM1, FC_OFDM2, FC_OFDM3}. First, time-interleaving approaches have been presented \cite{SC_OFDM1, SC_OFDM2, SC_OFDM3}. A stepped-carrier (SC) OFDM radar system achieves low ADC/DAC sampling rates by means of local oscillator (LO) frequency stepping \cite{SC_OFDM1}. The time-interleaved transmission of each sub-band at each LO stepping frequency allows this approach to reduce the baseband sampling rate according to the number of sub-bands. However, this scheme degrades the maximum unambiguous velocity of the OFDM radar due to an increase in the transmit symbol duration. 

Another approach to reducing the baseband sampling rate is to use a frequency-interleaving technique, referred to as a frequency comb (FC) OFDM \cite{FC_OFDM1, FC_OFDM2, FC_OFDM3}. FC-OFDM radar deploys a frequency comb generator that creates multiple LO carriers and mixes them with the baseband OFDM signal. In this way, it transmits the full-band signal simultaneously, mitigating the degradation of the maximum unambiguous velocity compared to a time-interleaving case. However, this method employs a subcarrier-interleaved sub-band signal, which limits the maximum unambiguous range of the OFDM radar. Furthermore, both time- and frequency-interleaving techniques not only incur additional hardware complexity but also cause phase discontinuities between the sub-bands due to the manipulation of the LO frequencies, which must be corrected before radar processing \cite{SC_OFDM4, SC_OFDM5, FC_OFDM3}. 

Although both time-and frequency-interleaving approaches have successfully achieved low baseband sampling rates while maintaining high range resolutions, the increased RF transceiver complexity is not preferable in a low-cost radar system. For low-cost and power-efficient radar systems, there have been efforts that use sub-Nyquist sampling of the radar signals in range as well as Doppler and spatial domains \cite{SNS1, SNS3, SNS4}. The main contributions of these systems are that the radar processing techniques employ a compressed sensing (CS) technique that recovers the full signals from sparsely undersampled signals obtained by random linear sampling. One example of the application of this approach to OFDM radar can be seen in \cite{SNS5}, which presented a frequency-agile sparse OFDM. This transmits randomly designed narrowband OFDM signals mixed according to certain LO frequency-hopping patterns. The sparse OFDM exploits CS to recover the sparse OFDM waveform, achieving a high range resolution without any range-Doppler ambiguities. However, the CS-based approach is still cumbersome and computationally complex, and its performance cannot be ensured for low signal-to-noise ratio (SNR) targets.  

Recently, Lang $et$ $al.$  presented a subcarrier aliasing (SA) OFDM radar system that takes advantage of sub-Nyquist sampling without an increase in the hardware or signal processing complexity \cite{SA_OFDM}. In SA-OFDM, designing a proper active subcarrier interval in the OFDM signal allows active subcarriers in each sub-band not to be overlapped in the undersampled signal. Although this involves simple hardware and radar signal processing compared to other approaches, its system performances in terms of the maximum unambiguous range and the range processing gain are severely degraded due to the reduced number of active subcarriers. Therefore, a simple solution is still required to reduce the ADC sampling rate without both degradation of the range and Doppler ambiguity.  
  
In this paper, we propose a sub-Nyquist sampling OFDM radar system that reduces the ADC sampling rate without any additional manipulations of the hardware or waveforms. This ensures a high range resolution and the maximum detectable range without ambiguities. Instead of using sparse non-uniform undersampling or a subcarrier-interleaved waveform, the proposed approach utilizes signal folding stemming from the uniform sub-Nyquist sampling. Signal demodulation of the sub-sampled signal enables unfolding of the folded signal to the full-band signal, which is accompanied by symbol-mismatch noise (SMN). The symbol-mismatch noise is then removed using the proposed SMN cancellation process, which is simple but effectively recovers the dynamic range of the range profiles obtained from the unfolded signal. The proposed scheme allows for a reduction of the ADC sampling rate at the slight expense of noise folding, which is prevailing and unavoidable during sub-Nyquist sampling without the need to reconfigure the system hardware.

This paper is organized as follows. Section \uppercase\expandafter{\romannumeral2} introduces the sub-Nyquist sampling OFDM radar system and corresponding signal demodulation method with an analysis of the SNR. In Section \uppercase\expandafter{\romannumeral3}, the symbol-mismatch noise cancellation process that improves the dynamic range of the SNS-OFDM radar system is proposed. Section \uppercase\expandafter{\romannumeral4} provides simulation results of the SNS-OFDM radar system. Measurement results and a discussion of the proposed work are presented in Section \uppercase\expandafter{\romannumeral5}. The conclusions are drawn in Section \uppercase\expandafter{\romannumeral6}.

$Notations$: Boldface variables with lower- and upper-case symbols represent vectors and matrices, respectively. $\textbf{A} \in \mathbb{C}^{N \times M}$ denotes a complex-valued ${N \times M}$ matrix $\textbf{A}$. $\textbf{A}_{k}$ is the $k^{th}$ submatrix of a row-wise block matrix $\textbf{A}$. $a_{ij}$ indicates an element of matrix $\textbf{A}$ at the $i^{th}$ row and the $j^{th}$ column. $({\cdot})^{T}$ and $({\cdot})^{H}$ represent the transpose and Hermitian transpose operators, respectively. ${\text{diag}({\textbf{a}})}$ denotes a diagonal matrix with diagonal entries of a vector $\textbf{a}$. $\odot$ and $\oslash$ represent the Hadamard (element-wise) product and division operator, respectively. $\mathbb{E}{[\cdot]}$ is the statistical expectation operator. 

\section{Sub-Nyquist Sampling OFDM Radar}
In this section, we introduce the signal model of the proposed sub-Nyquist sampling OFDM radar system. To simplify the description, a single-input single-output (SISO) radar system is assumed, which can be readily extended to a multiple-input multiple-output (MIMO) radar system. 
\subsection{Signal model of OFDM radar}
A conventional OFDM radar transmitter (TX) transmits multiple numbers of symbols $N_s$ within one frame. One OFDM symbol consists of $N_c$ subcarriers, which are modulated with certain constellation symbols. Here, let $\textbf{C}$ denote the transmitted symbol matrix. This can be expressed as

\begin{equation}
\textbf{C} = 
    \begin{bmatrix}
    c_{11} & c_{21} & \cdots & c_{N_{s}1} \\
    c_{12} & c_{22} & \cdots & c_{N_{s}2} \\
    \vdots & \vdots & \ddots & \vdots \\
    c_{1N_{c}} & c_{2N_{c}} & \cdots & c_{N_{s}N_{c}}
    \end{bmatrix}
    \in \mathbb{C}^{N_c \times N_s}.
    \label{E1}
\end{equation}
We exploit a random code matrix $\textbf{C}$, which means that the input data sequences used for OFDM symbol generation are random.

\begin{figure*}[t!]
    \centering
    {\includegraphics[scale=0.5]{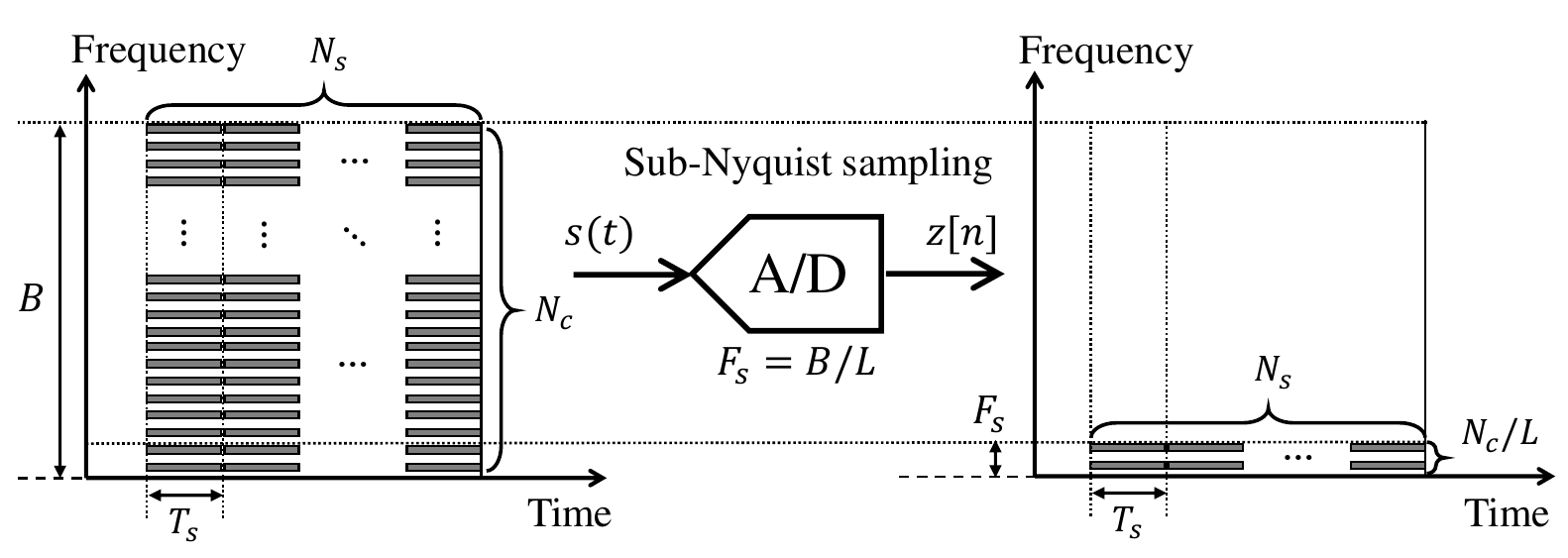}}
    \caption{Operation of sub-Nyquist sampling OFDM radar. Received signals with signal bandwidth $B$ are sampled by an ADC with a sampling rate of $B/L$ .}
    \label{f1}
\end{figure*}

\begin{figure}[t!]
    \centering
    {\includegraphics[scale=0.4]{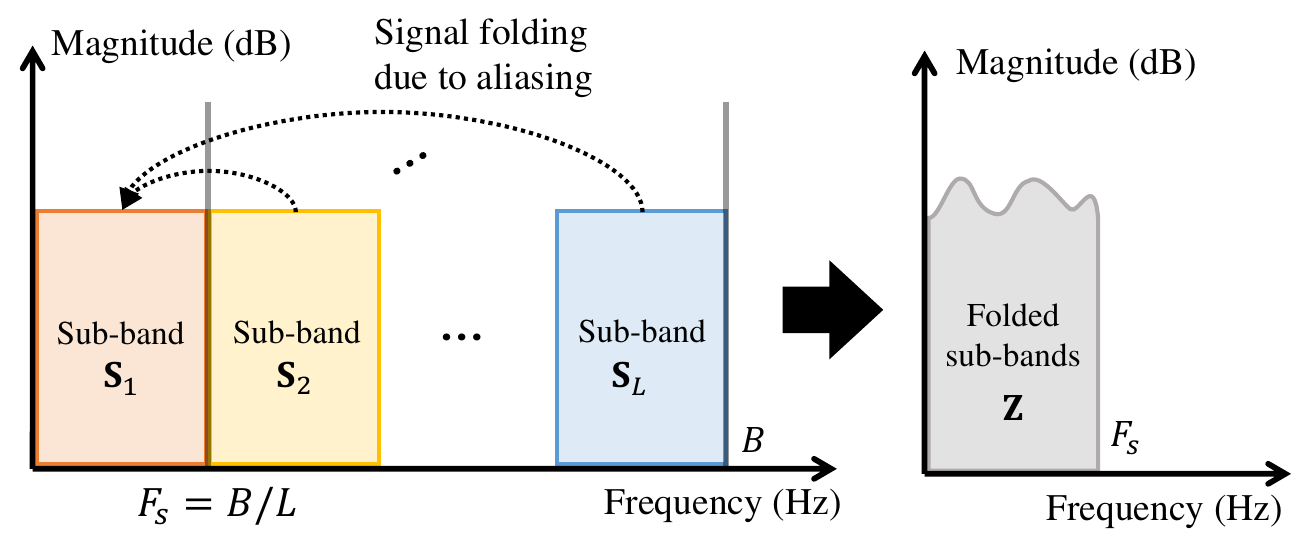}}
    \caption{Spectral representation of OFDM signal folding due to sub-Nyquist sampling.}
    \label{f2}
\end{figure}

After the transmitted signal is reflected by targets, the receiver (RX) receives scattered signals from all of the targets. The signal of each target produces a round-trip time-delay $\tau$ and a Doppler frequency shift $\omega$ corresponding to its range and relative velocity to the radar, respectively. Assuming that the target range and velocity are constant within one radar frame, the round-trip time delay induces a linear phase shift along the subcarrier axis and the Doppler frequency do so along the symbol axis. Therefore, the range steering vector $\textbf{r}(\tau)$ and Doppler steering vector $\textbf{v}(\omega)$ of each target are given as 
\begin{equation}
    \textbf{r}(\tau) = 
    \begin{bmatrix}
        e^{-j2 \pi \Delta f \tau}, e^{-j2 \pi 2\Delta f \tau}, \cdots, e^{-j2 \pi N_c\Delta f \tau}
    \end{bmatrix}^{T}\in \mathbb{C}^{N_c \times 1},
    \label{E2}
\end{equation}
\begin{equation}
    \textbf{v}(\omega) = 
    \begin{bmatrix}
        e^{j \omega T_s}, e^{j 2\omega T_s},  \cdots, e^{j N_s\omega T_s}
    \end{bmatrix}^{T} \in \mathbb{C}^{N_s \times 1},
    \label{E3}
\end{equation}
where $\Delta f = B/N_s$ denotes the subcarrier spacing of the OFDM signal, and $T_s = 1/\Delta f + T_{cp}$ is the OFDM symbol duration with the time duration $T_{cp}$ of a cyclic prefix. Given that the TX signal is reflected from $K$ number of targets, the target information matrices are defined as shown below.
\begin{equation}
    \textbf{R} = 
    \begin{bmatrix}
        \textbf{r}(\tau_1) & \textbf{r}(\tau_2) & \cdots & \textbf{r}(\tau_K)
    \end{bmatrix}^{T}\in \mathbb{C}^{N_c \times K},
    \label{E4}
\end{equation}
\begin{equation}
    \textbf{V} = 
    \begin{bmatrix}
        \textbf{v}(\omega_1) & \textbf{v}(\omega_2) & \cdots & \textbf{v}(\omega_K)
    \end{bmatrix}^{T} \in \mathbb{C}^{N_s \times K},
    \label{E5}
\end{equation}
\begin{equation}
    \textbf{A} = 
    \text{diag}(\alpha_1,\alpha_2, \cdots, \alpha_K) 
    \in \mathbb{C}^{K \times K},
    \label{E6}
\end{equation}
where $\textbf{R}$ represents the range matrix and $\textbf{V}$ denotes the Doppler matrix. Also, $\textbf{R}$ contains the complex amplitude of each target, including the target radar cross-section (RCS).

Generally, the received OFDM radar signal is converted to a frequency-domain signal by means of the fast Fourier transform (FFT) before signal demodulation is conducted. The frequency-domain representation of the received signal $\textbf{S}$ can be expressed as 
\begin{equation}
    \textbf{S} = \textbf{C} \odot \textbf{X} + \textbf{W} \in \mathbb{C}^{N_c \times N_s}, 
    \label{E7}
\end{equation}
where
\begin{equation}
    \textbf{X} = \textbf{R}\textbf{A}\textbf{V}^{T}.
    \label{E8}
\end{equation}
$\textbf{W}$ is the complex white Gaussian noise with i.i.d $\mathcal{CN}(0,\,\sigma^{2})$ entries. To reconstruct $\textbf{S}$ fully without signal aliasing, the ADC sampling rate must be larger than the signal bandwidth $B$ in case both in-phase ($I$) and quadrature ($Q$) signals are sampled. In the conventional OFDM radar system, 
the radar parameters are estimated after modulation-symbol-based demodulation over $\textbf{S}$.

\subsection{Sub-Nyquist sampling (SNS) OFDM radar}
The proposed sub-Nyquist sampling (SNS) OFDM radar system exploits an ADC with uniform sub-Nyquist sampling rate $B/L$ to sample received intermediate-frequency (IF) signals, of which the signal bandwidth is $B$. Fig. \ref{f1} shows the operation of the SNS-OFDM radar system, which has a waveform structure identical to that described in the previous section. The sub-sampling ratio $L$ can be any positive proper divisor of the number of subcarriers $N_c$. As we use sub-Nyquist sampling, the sampled signal is a folded signal with $L$ sub-bands fully overlapped within the bandwidth of $B/L$ due to aliasing.

From the frequency-domain representation of the received signal (\ref{E7}), $\textbf{S}$, $\textbf{C}$, $\textbf{X}$, and $\textbf{W}$ can be partitioned into $L$ submatrices as
\begin{equation}
    \begin{bmatrix}
    \textbf{S}_1 \\ \textbf{S}_2 \\ \vdots \\ \textbf{S}_{L}
    \end{bmatrix}
    = \begin{bmatrix}
    \textbf{C}_1 \\ \textbf{C}_2 \\ \vdots \\ \textbf{C}_{L}
    \end{bmatrix}
    \odot \begin{bmatrix}
    \textbf{X}_1 \\ \textbf{X}_2 \\ \vdots \\ \textbf{X}_{L}
    \end{bmatrix}
    + \begin{bmatrix}
    \textbf{W}_1 \\ \textbf{W}_2 \\ \vdots \\ \textbf{W}_{L}
    \end{bmatrix},
    \label{E10}
\end{equation}
where $\textbf{S}_k$, $\textbf{C}_k$, $\textbf{X}_k$, and $\textbf{W}_k$ $\in \mathbb{C}^{(N_c/L) \times N_s}$ for $k = 1, 2, ... , L$. From (\ref{E10}), the sub-sampled signal is equivalently expressed as the summation of all sub-band signals in the frequency domain. Fig. \ref{f2} shows a spectral representation of signal folding as it pertains to the proposed SNS-OFDM radar system. Let $\textbf{Z}$ be the sub-sampled signal that is represented in the frequency domain, which can then be written as follows:

\begin{align}
    \textbf{Z} & = \textbf{S}_1 + \textbf{S}_2 + \cdots + \textbf{S}_L \\
    & = {\sum_{k=1}^{L}} \,(\textbf{C}_k \odot \textbf{X}_k) + {\sum_{k=1}^{L}} \, \textbf{W}_k.
    \label{E11}
\end{align}

\begin{figure*}[t!]
    \centering
    {\includegraphics[scale=0.53]{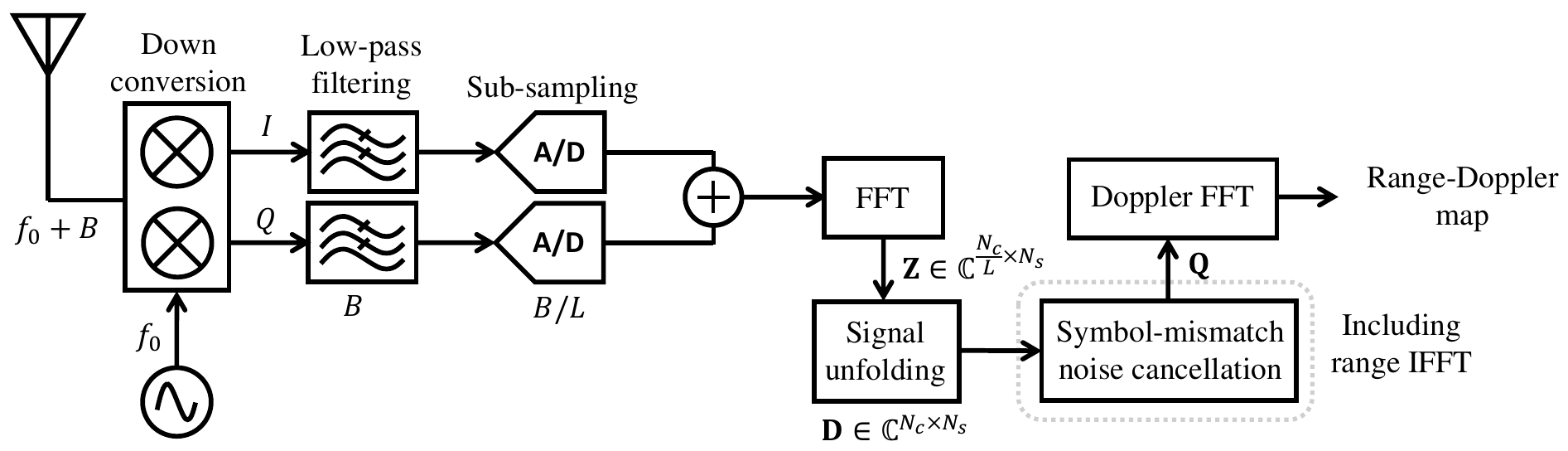}}
    \caption{Block diagram of the sub-Nyquist sampling OFDM radar receiver and corresponding signal processing chain.}
    \label{f3}
\end{figure*}
\begin{figure}[t!]
    \centering
    {\includegraphics[scale=0.4]{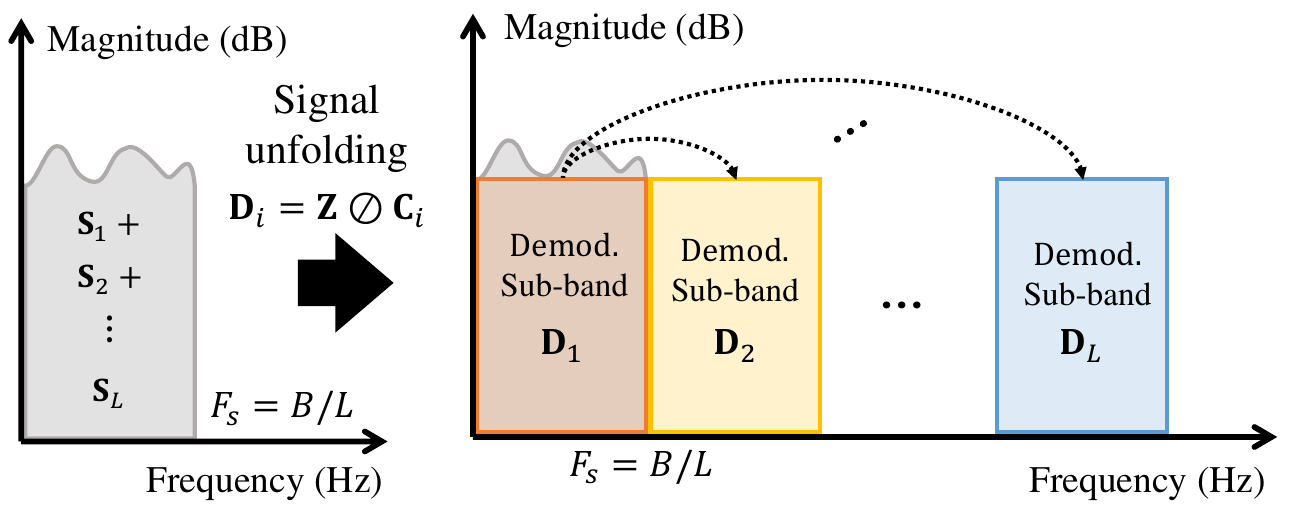}}
    \caption{Spectral representation of the signal-unfolding process in sub-Nyquist sampling OFDM radar.}
    \label{f4}
\end{figure}

Because the bandwidth of the folded signal is reduced to $B/L$, the range resolution is also degraded to the extent of $L$ before signal demodulation processing is conducted. It should be noted that the proposed SNS-OFDM radar allows $L$ sub-band signals to be overlapped into the same frequency band. This differs from the subcarrier-aliasing OFDM radar system reported in earlier work \cite{SA_OFDM}, which avoids signal folding by exploiting subcarrier-interleaved waveforms. The proposed sub-Nyquist sampling process can be utilized with any type of OFDM waveform $\textbf{C}$ generated by random data sequences. 

With the signal form of (\ref{E11}), a conventional demodulation scheme for OFDM radar cannot give the full range resolution corresponding to the signal bandwidth $B$. Nevertheless, the sub-sampled signal can be fully recovered to the full-band signal by using the transmitted symbol matrix $\textbf{C}$. The key idea in SNS-OFDM radar is that each sub-band can be separated from the folded signal by using prior knowledge of the transmitted OFDM symbols. Details of this signal demodulation method are discussed in the following section.  

\subsection{Signal demodulation}
In this section, we describe the signal demodulation method used in the proposed SNS-OFDM radar system, referred to as the signal-unfolding process. Fig. \ref{f3} shows a block diagram of the proposed SNS-OFDM radar receiver and the corresponding overall signal processing chain. After sampling the received signal, the frequency-domain sub-sampled signal $\textbf{Z}$ is obtained by performing FFTs. Because the sub-sampled signal is now folded into one sub-band with the bandwidth of $B/L$, it must initially be unfolded to a full-band signal. From the folded signal, the $i^{th}$ sub-band can be demodulated by using the $i^{th}$ sub-band symbol matrix $\textbf{C}_i$. The demodulated $i^{th}$ sub-band signal $\textbf{D}_i \in \mathbb{C}^{(N_c/L) \times N_s}$ can be derived as
\begin{align}    
    \textbf{D}_i & = \textbf{Z} \oslash \textbf{C}_i \\
    & = \left({\sum_{k=1}^{L}} \, ({\textbf{C}_k \odot \textbf{X}_k) + {\sum_{k=1}^{L}} \, \textbf{W}_k}\right) \oslash \textbf{C}_i\\
    & = \textbf{X}_i + {\sum_{k = 1, k \ne i}^{L}} \, \left(\textbf{X}_k \odot (\textbf{C}_k \oslash \textbf{C}_i)\right) + {\sum_{k = 1}^{L}} \, \textbf{W}_k \oslash \textbf{C}_i.    
    \label{E12}
\end{align}
Let $\textbf{D}$ denote the full-band signal, which can be devised as an augmented matrix of demodulated sub-bands $\textbf{D}_i$. Then, $\textbf{D}$ can be expressed as 
\begin{align}
        \textbf{D} & = 
        \begin{bmatrix}
            \textbf{D}_1^{T} & \textbf{D}_2^{T} & \cdots & \textbf{D}_\textit{L}^{T} 
        \end{bmatrix}^{T} \in \mathbb{C}^{N_c \times N_s} \\
        & = \begin{bmatrix}
        \textbf{X}_1 \\ \textbf{X}_2 \\ \vdots \\ \textbf{X}_\textit{L}
        \end{bmatrix}+ 
         \begin{bmatrix}
        {\sum_{k\ne 1}} \, \textbf{X}_k \odot (\textbf{C}_k \oslash \textbf{C}_1) \\ {\sum_{k\ne 2}} \, \textbf{X}_k \odot (\textbf{C}_k \oslash \textbf{C}_2) \\ \vdots \\ {\sum_{k\ne L}} \, \textbf{X}_k \odot (\textbf{C}_k \oslash \textbf{C}_L)
         \end{bmatrix} + 
        \begin{bmatrix}
        {\sum_{k=1}^{L}} \, \textbf{W}_k \oslash \textbf{C}_1 \\
        {\sum_{k=1}^{L}} \, \textbf{W}_k \oslash \textbf{C}_2 \\ \vdots
        \\ {\sum_{k=1}^{L}} \, \textbf{W}_k \oslash \textbf{C}_L
        \end{bmatrix} \\ 
        & = \textbf{X} + \textbf{Y} + \textbf{W}_F, 
    \label{E13}
\end{align}
where $\textbf{W}_F$ represents the folded white Gaussian noise due to sub-Nyquist sampling. In (\ref{E13}), it is shown that a folded signal due to sub-Nyquist sampling can be expanded to a full-band signal, which includes the target information matrix $\textbf{X}$. At this point, we can take advantage of the full range resolution equivalent to that of the original IF signal bandwidth $B$. This is termed the signal-unfolding process, as shown in Fig. \ref{f4}.

Although the signal-unfolding process provides an improved range resolution, it involves unexpected residual noise $\textbf{Y}$ in (\ref{E13}). This is the summation of the interference signals from other sub-bands. They are multiplied by each submatrix of the block symbol matrix during the signal-unfolding process. Because the transmitted modulation symbol $\textbf{C}$ is generated by random data sequences, the submatrices of the block symbol matrix $\textbf{C}_i$ are uncorrelated with each other for all sub-bands. Therefore, $\textbf{C}_k \oslash \textbf{C}_i$ in $\textbf{Y}$ acts as random noise for all $k = 1, 2, ... , L, k\ne i$. Here, we refer to the term $\textbf{Y}$ as the symbol-mismatch noise (SMN). This symbol-mismatch noise increases the integrated sidelobe level (ISL) in the range profile, which can be obtained through inverse FFT (IFFT) over the unfolded signal $\textbf{D}$. It not only degrades the detected target SNR but also causes weak targets to be submerged by the increased ISL of a strong target. In the following section, we conduct a precise analysis of the effect of the symbol-mismatch noise and noise folding in SNS-OFDM radar.  

\subsection{SNR analysis}
The signal-folding and -unfolding processes in the SNS-OFDM radar system introduce additional noise compared to the conventional OFDM radar system. Here, we explore the signal-to-noise ratio (SNR) of detected targets from the unfolded signal. First, the noise folding effect causes an increased noise level of the sub-sampled signal in the SNS-OFDM radar system, which is represented as $\textbf{W}_F$ in (\ref{E13}). Compared to the Nyquist sampling scheme, a system using sub-sampling inevitably undergoes noise folding. Because $\textbf{W}_F$ is the summation of the complex white Gaussian noise of $L$ sub-bands, it is equivalent to the complex white Gaussian noise with i.i.d $\mathcal{CN}(0,\,\sigma^{2}L)$ entries. The folded noise level increases in proportion with the sub-sampling ratio $L$ in SNS-OFDM radar.
Given that the transmitted OFDM symbol is generated by completely random data sequences, a second additional noise source is the symbol-mismatch noise $\textbf{Y}$ generated by signal folding. As shown in (\ref{E13}), the SMN is determined by not only the sub-sampling ratio $L$ but also by the amplitudes of the target signals. Using (\ref{E6}), (\ref{E8}), and (\ref{E13}), a generic form of the detected SNR in a single range profile can be derived as follow:      
\begin{align}
        \textrm{SNR}_\textrm{o} & = 10\log_{10} \left(\frac{\mathbb{E}[\textbf{X}\textbf{X}^H]}{\mathbb{E}[\textbf{Y}\textbf{Y}^H] + \sigma^2 L}\right) + 10\log_{10} {N_c}\\
        & = 10\log_{10} \left(\frac{N_c{\sum_{k=1}^{K}} \, |\alpha_k|^2}{(L-1){\sum_{k=1}^{K}} \, |\alpha_k|^2 + \sigma^2 L}\right).
    \label{E14}
\end{align}
\begin{figure}[t!]
    \centering
    {\includegraphics[scale=0.5]{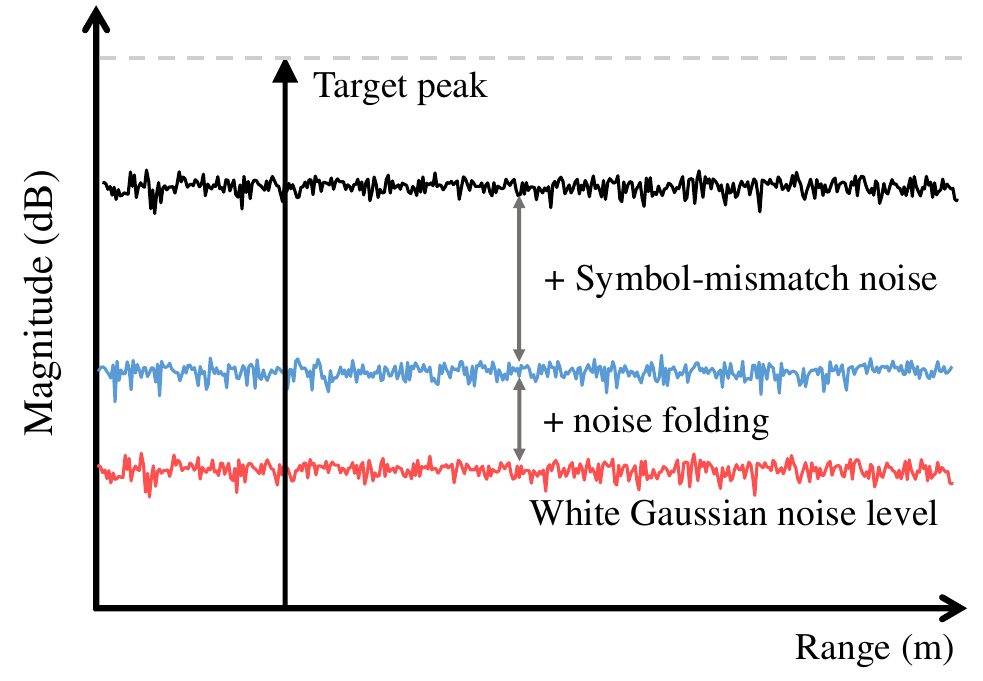}}
    \caption{Noise level in the range profile of sub-Nyquist sampling OFDM radar.}
    \label{f5}
\end{figure}

The signal power in (\ref{E14}) takes advantage of the range processing gain to the extent of $10\log_{10}{N_c}$. On the other hand, the SMNs in $(L-1)$ sub-bands are non-coherently combined because they can be regarded as independent and identically distributed random variables. A graphical illustration of the detected SNR of the proposed SNS-OFDM radar system is presented in Fig. \ref{f5}. Assuming that a single strong target exists, the detected SNR of the target (\ref{E14}) is determined by only the SMN. This is simply reduced to 

\begin{equation}
    \begin{aligned}
        \textrm{SNR}_\textrm{o}  \approx 10\log_{10} \left(\frac{N_c}{L-1}\right). 
    \end{aligned}
    \label{E15}
\end{equation}.
This demonstrates that the theoretical SNR of the SNS-OFDM radar system is limited according to the number of subcarriers and the sub-sampling ratio. This equals the number of subcarriers in each sub-band. In the presence of SMN, it is clear that the detection performance is severely degraded compared to that in the conventional OFDM radar system. Therefore, a solution to cancel out the SMN is required in the proposed SNS-OFDM radar system. In the following section, we investigate a symbol-mismatch noise cancellation (SMNC) process that recovers the unfolded signal to one without SMN.    
\begin{figure}[t!]
    \centering
    {\includegraphics[scale=0.42]{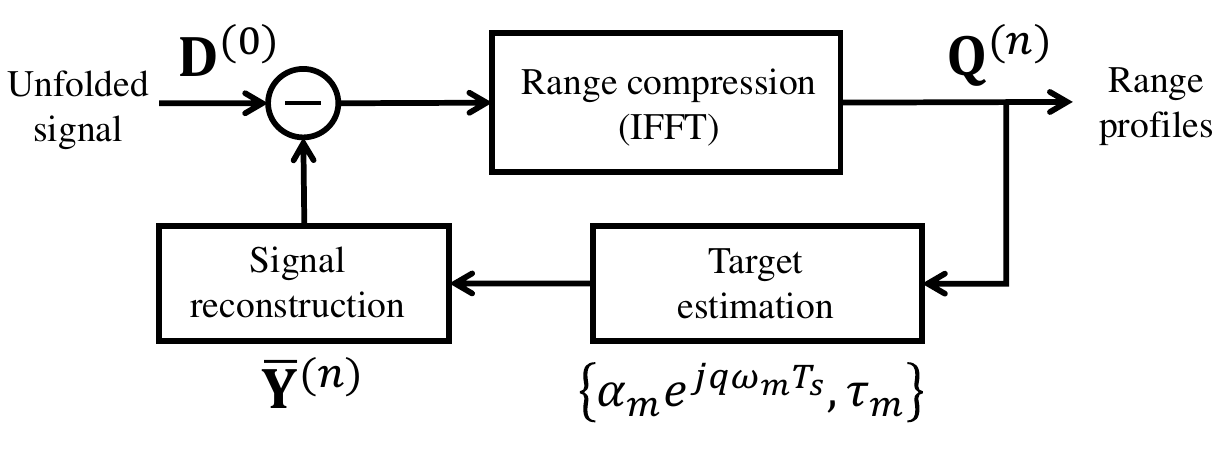}}
    \caption{Signal processing flow chart of symbol-mismatch noise cancellation (SMNC) in sub-Nyquist sampling OFDM radar.}
    \label{f6}
\end{figure}
\section{Symbol-Mismatch Noise Cancellation (SMNC) in Sub-Nyquist Sampling OFDM Radar}
In this section, we explore a method to mitigate the symbol-mismatch noise additionally introduced in SNS-OFDM radar. After the the signal-unfolding process, the SMN severely increases the ISL in the range profile, submerging weak targets below the ISL of a strong target. Therefore, it is necessary for the SMN to be canceled out to ensure that the proposed SNS-OFDM radar system has detection performance comparable to that of conventional OFDM radar using Nyquist sampling. 

Given that the number of total subcarriers is much greater than the number of sub-bands, strong targets with $|\alpha_k|^2 \gg \sigma^2 L$ are always detectable because their SNRs are always larger than $0$ dB, as shown in (\ref{E15}). Therefore, we exploit the detected strong targets and transmitted OFDM symbols to cancel out the SMN in SNS-OFDM radar. The overall signal processing flow of the proposed SMNC technique is described in Fig. \ref{f6}. Here, we exploit an iterative approach of target detection and interfering signal reconstruction in the range profiles of the SNS-OFDM radar system. Each term indicated at each node in Fig. \ref{f6} represents the output of the $n^{th}$ iteration.

Let the initial unfolded signal be denoted as $\textbf{D}^{(0)}$. At the first iteration stage, to obtain the range profile from the unfolded signal, we utilize ${N}_{c}$-points IFFT over each column of $\textbf{D}^{(0)}$, as 
\begin{equation}
        \textbf{Q}^{(0)}  = \textrm{IFFT}{[\textbf{D}^{(0)}]}. 
    \label{E16}
\end{equation}
To estimate the targets from $\textbf{Q}^{(0)}$, we utilize a cell-average constant false alarm rate (CA-CFAR) detector to non-coherently combine $N_s$ range profiles. Therefore, only one detection process is required per iteration stage. Assume that $M_{(0)}$ targets among $K$ targets are detected at the first detection, their parameters $[\alpha_m e^{j q \omega_{m} T_s}, and \tau_{m}]$ for $m = 1, 2, ..., M_{(0)}$ can be specified by extracting the signals at their ranges in the $q^{th}$ range profile obtained from the $q^{th}$ column of $\textbf{Q}^{(0)}$. It is important to note that the estimated target $\overline{\textbf{X}}^{(0)}$ also includes the effect of the SMN because the selected range bins contain not only the target parameters but also the SMNs that spread throughout the range profile. However, the level is negligible due to iterative processing, which reduces the SMN power at a certain target range bin by a factor of $(L-1)/N_c$ per iteration.
\begin{algorithm}[t!]
     \caption{Symbol-mismatch noise cancellation}
     \begin{algorithmic}[1]
     \renewcommand{\algorithmicrequire}{\textbf{Input:}}
     \renewcommand{\algorithmicensure}{\textbf{Output:}}
     \REQUIRE Unfolded signal $\textbf{D}$ $\in \mathbb{C}^{N_c \times N_s}$, and $\textbf{C}$ $\in \mathbb{C}^{N_c \times N_s}$
     \ENSURE  Range profile $\textbf{Q} \in \mathbb{C}^{N_c \times N_s}$ 
     \\ \textit{Initialisation} : $\textbf{D}^{(0)} = \textbf{D}$, $n = 1$, $\mu_{-1} = 0$
        \STATE $\textbf{Q}^{(0)}  = \textrm{IFFT}{[\textbf{D}^{(0)}]}$ 
        \STATE Calculate the initial sidelobe level $\mu_{0}$ from $\textbf{Q}^{(0)}$
          
      \WHILE{$|\mu_{n-1} - \mu_{n-2}| > \epsilon$}
          \STATE Estimate target parameters $\overline{\textbf{X}}^{(n-1)}$ from $\textbf{Q}^{(n-1)}$
          \STATE Reconstruct $\overline{\textbf{Y}}^{(n-1)}$ using $\overline{\textbf{X}}^{(n-1)}$ and $\textbf{C}$
          \STATE $\textbf{D}^{(n)} = \textbf{D}^{(0)} - \overline{\textbf{Y}}^{(n-1)}$
          \STATE $\textbf{Q}^{(n)}  = \textrm{IFFT}{[\textbf{D}^{(n)}]}$
          \STATE Calculate the sidelobe level $\mu_{n}$ from $\textbf{Q}^{(n)}$
          \STATE $n$ = $n + 1$
     \ENDWHILE
     \end{algorithmic}
     \label{A}
 \end{algorithm} 
 
With the estimated target signals $\overline{\textbf{X}}^{(0)}$ and known transmitted symbols $\textbf{C}$, the symbol-mismatch noise imposed only by the detected targets can be reconstructed as shown below. 
\begin{equation}
    \begin{aligned}
        \overline{\textbf{Y}}^{(0)} & = 
         \begin{bmatrix}
        {\sum_{k\ne 1}} \, \overline{\textbf{X}}^{(0)}_k \odot (\textbf{C}_k \oslash \textbf{C}_1) \\ {\sum_{k\ne 2}} \, \overline{\textbf{X}}^{(0)}_k \odot (\textbf{C}_k \oslash \textbf{C}_2) \\ \vdots \\ {\sum_{k\ne L}} \, \overline{\textbf{X}}^{(0)}_k \odot (\textbf{C}_k \oslash \textbf{C}_L)
         \end{bmatrix}. 
    \end{aligned}
    \label{E17}
\end{equation}
The reconstructed signal is subtracted from the initial unfolded signal, which updates the unfolded signal, as follows:
\begin{align}
        \textbf{D}^{(1)} & = \textbf{D}^{(0)} - \overline{\textbf{Y}}^{(0)} \\
        & = \textbf{X} - (\textbf{Y} - \overline{\textbf{Y}}^{(0)}) + \textbf{W}_{F}.
    \label{E18}
\end{align}

Because the SMN depends on the power of the targets, the SMNC term $(\textbf{Y} - \overline{\textbf{Y}}^{(0)})$ reduces the sidelobe level in the range profile. Also, it allows for the detection of weak targets that are not detected at the first detection stage. Again, we can perform the same detection and reconstruction procedure at the next iteration step. As the number of steps increases, more targets are detected in the range profile. For the $n^{th}$ iteration, $M_{(n)}$ number of targets are exploited to reconstruct the SMN $\overline{\textbf{Y}}^{(n-1)}$. Consequently, the unfolded signal after $n$ number of iterations can be expressed as follows:    
\begin{align}
        \textbf{D}^{(n)} & = \textbf{D}^{(0)} - \overline{\textbf{Y}}^{(n-1)} \\
        & = \textbf{X} - (\textbf{Y} - \overline{\textbf{Y}}^{(n-1)}) + \textbf{W}_{F}.
    \label{E19}
\end{align}

The proposed method repeats this procedure until the difference in the sidelobe levels between consecutive iterations becomes smaller than a certain threshold such that $|\mu_{n} - \mu_{n-1}| < \epsilon$. This convergence condition is equivalent to that which the sidelobe level reaches at the folded noise level $\sigma^2 L$, which means that all symbol-mismatch noise sources are removed. Given that SMNC processing converges after $N$ number of iterations, the requirement of all relationships in the following descriptions are satisfied.
\begin{align}
        M_{(N)} &= K, \\
        \overline{\textbf{Y}}^{(N-1)} & = \textbf{Y},\\
        \textbf{D}^{(N)} & = \textbf{X} + \textbf{W}_{F}.
    \label{E20}
\end{align}
The result of the proposed SMNC processing $\textbf{Q}^{(N)}$ is the range profile without the SMN, which is then processed by Doppler FFT. In (\ref{E20}), the output only includes the folded noise due to aliasing. Finally, we can determine the SNR after canceling the SMN, as follows:
\begin{equation}
    \begin{aligned}
        \textrm{SNR}_{N}  = 10\log_{10} \left(\frac{N_c{\sum_{k=1}^{K}} \, |\alpha_k|^2}{\sigma^2 L}\right). 
    \end{aligned}
    \label{E21}
\end{equation}

Although the proposed SMNC slightly increases the computational complexity compared to that of the conventional OFDM radar system, only a few iterations are required in practice because the proposed method can utilize all of the detected targets in each iteration to reconstruct the SMN. The number of iterations required for convergence depends on the folded noise level, not on the number of targets, $K$. The proposed algorithm is summarized in Algorithm \ref{A}. In the following section, simulation results with SNS-OFDM radar and SMNC processing are provided to demonstrate the effectiveness of the proposed radar system.     

\section{Simulation Results of Sub-Nyquist Sampling OFDM Radar}
\begin{table}[t!]
\centering
\fontsize{8.3}{14}\selectfont
\caption{OFDM radar parameters used in the simulations}
\label{t1}
\begin{tabular}{>{\centering\arraybackslash} m{17em} | >{\centering\arraybackslash} m{10em}}
    \toprule
    Parameters    & Values\\ \hline
    \midrule
    Number of subcarriers, $N_\textit{c}$         & 2048         \\ \hline
    Subcarrier spacing, $\Delta f$               & 488.28 kHz     \\ \hline
    Total signal bandwidth, $B$                       & 1 GHz     \\ \hline
    CP duration,  $T_\textit{CP}$                   & 0.512 $\mu$s \\ \hline
    OFDM symbol duration w/o CP, $1/{\Delta f}$          & 2.048 $\mu$s    \\ \hline
    OFDM symbol duration w/ CP, $T_\textit{s}$ & 2.56 $\mu$s \\ \hline
    Carrier frequency, $f_\textit{c}$               & 77 GHz  \\ \hline
    ADC sampling rate, ${F}_{s}$               & 1000, 500, 250, 125, 62.5, and 31.25 MSamples/s  \\
    \bottomrule
\end{tabular}
\end{table}

\begin{figure}[t!]
    \centering
    \subfloat[]{\includegraphics[scale=0.75]{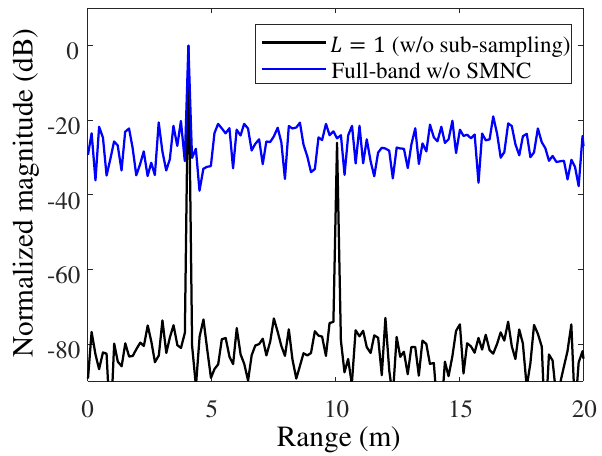}}
    
    \centering
    
    \subfloat[]{\includegraphics[scale=0.75]{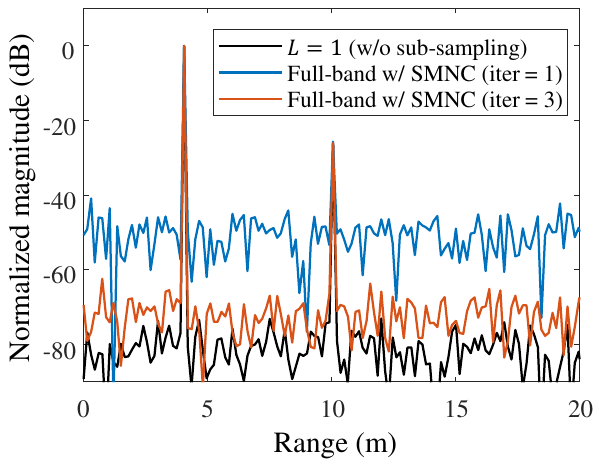}}
    \caption{Exemplary simulation results of sub-Nyquist sampling OFDM radar. Range profiles (a) without symbol-mismatch noise cancellation, and (b) those with symbol-mismatch noise cancellation.}
    \label{f7}
\end{figure}
\begin{figure}[t!]
    \centering
    \subfloat[]{\includegraphics[scale=0.74]{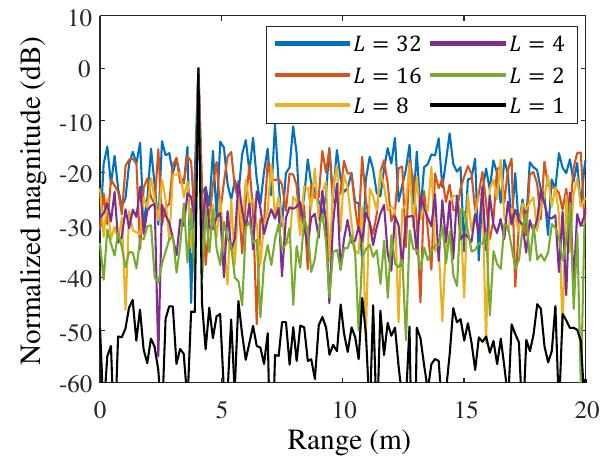}}
    
    \centering
    
    \subfloat[]{\includegraphics[scale=0.74]{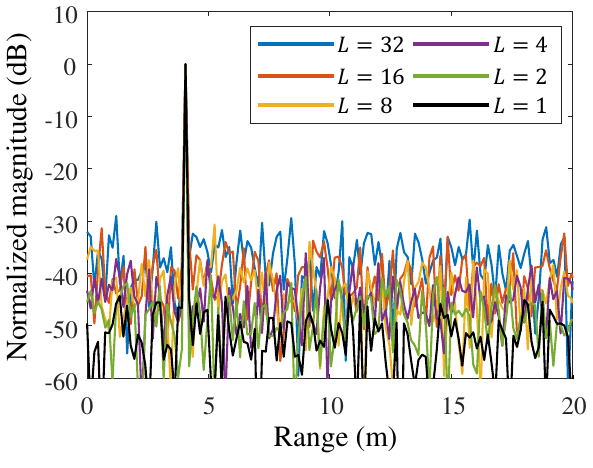}}
    \caption{Simulated range profiles obtained from unfolded full-band signals with various sub-sampling ratios $L$ (a) without symbol-mismatch noise cancellation, and (b) those with symbol-mismatch noise cancellation.}
    \label{f8}
\end{figure}
To investigate the feasibility of the proposed sub-Nyquist sampling OFDM radar system, simulations with various sub-sampling ratios are conducted. Also, the effectiveness of the SMNC technique in SNS-OFDM radar is validated through simulations. Table \ref{t1} shows the OFDM radar parameters used in the simulation. The overall bandwidth of the OFDM symbol is 1 GHz, and it consists of 2048 subcarriers. This gives a subcarrier spacing of 488.28 kHz and a symbol duration of 2.048 $\mu$s, except for the CP. Here, the length of the CP is one quarter of that of one OFDM symbol, which corresponds to 0.512 $\mu$s. Thus, the total OFDM symbol duration including the CP is 2.56 $\mu$s. The range resolution of the simulated OFDM radar is 0.15 m. In this case, we only evaluate a single OFDM symbol to verify the effect of sub-Nyquist sampling.

\subsection{Exemplary simulation with two targets}

First, we undertake an exemplary simulation to show the step-by-step results of the proposed SMNC technique in SNS-OFDM radar. The noise power of this simulation is set to -80 dBm. Two targets are located at ranges of 4 m and 10 m, of which corresponding power levels are 0 dBm and -26 dBm. A sub-sampling ratio $L = 8$ is adopted, which means that the sampling rate of both ADCs for $I$ and $Q$ channels equals 125 MSamples/s. 

Fig. \ref{f7}(a) shows the range profiles of the full-band signal obtained from the unfolding process. To compare the result, the range profile of the method with Nyquist sampling ($L = 1$, $F_{s} = 1$ GHz) is also shown. In the range profile of the SNS-OFDM radar system, only one target located at 4 m is detected; the second one is submerged due to symbol-mismatch noise. On the other hand, the range profile with Nyquist sampling has two simulated targets. As analyzed previously, the detected target SNR follows the equation (\ref{E15}), which results in an ISL of -24.66 dB. Because the signal power of the second target is weaker than that of the ISL, it cannot be seen in the range profile of the SNS-OFDM radar system. 

To resolve this problem, we apply the proposed SMNC technique to the full-band signal. Fig. \ref{f7}(b) presents the range profile after each iteration of SMNC processing. At the first iteration stage, the detected target at 4 m is exploited to reconstruct the symbol-mismatch noise. By subtracting the reconstructed SMN from the unfolded signal, the SMN due to the first target is removed, and the second target can be seen after the first iteration. At this point, both targets are taken into account to reconstruct the SMN. After three iterations, the output of SMNC processing yields an ISL of -72.11 dB, which is much lower than that obtained before SMNC processing. We set $\epsilon = 0.5$ dB as a convergence condition for the SMNC algorithm. Here, we can check that the sidelobe level of the SMNC output approaches the theoretical folded noise level, which is computed as -80 dBm + $10\log_{10} {8}$ dB = -71 dBm. Consequently, this results confirm that the proposed SMNC technique causes SNS-OFDM radar to be degraded only by additive folded noise only to the extent of $10\log_{10} {L}$ compared to the conventional Nyquist sampling OFDM radar system.

\subsection{Effect of the sub-sampling ratio, $L$}
The proposed SNS-OFDM radar system is also simulated with various sub-sampling ratios, $L$. Here, we assume that one target exists at a range of 4 m with power of 0 dBm. The noise power is set to -50 dBm. Fig. \ref{f8}(a) shows the range profiles from the unfolded full-band signal with various sub-sampling ratios of $L$ = 1, 2, 4, 8, 16, and 32, corresponding to ADC sampling rates of $F_{s}$ = 1000, 500, 250, 125, 62.5, and 31.25 MSamples/s. The sub-sampling ratio $L = 1$ indicates Nyquist sampling, which exploits the same ADC sampling rate with a total signal bandwidth of 1 GHz. Compared to the range profile with Nyquist sampling, those of the SNS-OFDM radar system have higher sidelobe levels, akin to findings in (\ref{E15}). As the sub-sampling ratio $L$ increases, the sidelobe level also increases up to $10\log_{10} {(L-1)}$ on the dB scale. On the other hand, SMNC processing enhances the sidelobe level, as shown in Fig. \ref{f8}(b). At this point, it is determined as in the equation (\ref{E21}) without the effect of SMN. Nevertheless, the sidelobe level of the SMNC processing output also increases proportionally as $10\log_{10} {L}$ on the dB scale. For both results with and without SMNC processing, the penalty of the 3 dB sidelobe level is accompanied by the halving of the ADC sampling rate. 

\subsection{Doppler tolerance of sub-Nyquist sampling}
\begin{figure}[t!]
    \centering
    {\includegraphics[scale=0.75]{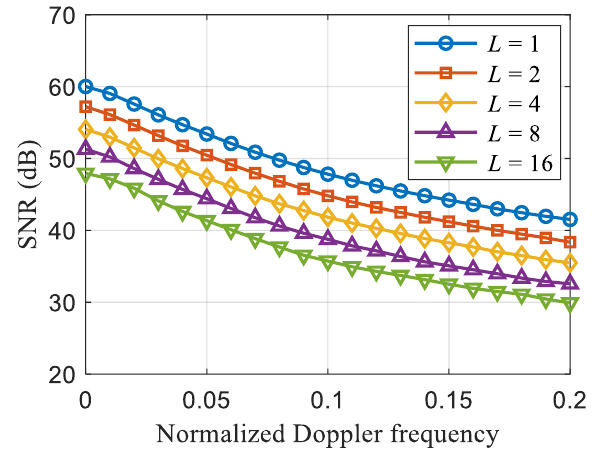}}
    \caption{Signal-to-noise ratio versus normalized Doppler frequency after symbol-mismatch noise cancellation in sub-Nyquist sampling OFDM radar.}
    \label{f9}
\end{figure}
Due to the target velocity, the received OFDM signal is marred by inter-carrier interference (ICI), which severely degrades the target SNR \cite{ICI}. To validate the performance of the proposed SNS-OFDM radar system with the velocity of the target, we simulate a moving target with various velocities. Here, we assume that the noise power is -60 dBm and that the target power is 0 dBm. Also, various sub-sampling ratios are tested with conditions identical to those in the simulation. The normalized Doppler frequency $\frac{\omega}{2\pi\Delta{f}}$ is set to range from 0 to 0.2.

Fig. \ref{f9} shows the simulated SNR of SNS-OFDM radar at the normalized Doppler frequency after SMNC processing. As expected, the SNR of the target is degraded as its velocity increases for all sub-sampling ratios. When the sub-sampling ratio increases to result in a lower ADC sampling rate, the SNR also decreases. However, there is no extra penalty from ICI in the SNS-OFDM radar case, of which the SNR is only affected by the sub-sampling ratio. Therefore, only the increased noise due to noise folding to the extent of $10\log_{10} {L}$ is a factor. 

\section{Measurement Results of Sub-Nyquist Sampling OFDM Radar}  
\begin{table}[t!]
\centering
\fontsize{8.3}{14}\selectfont
\caption{OFDM radar parameters for the measurements}
\label{t2}
\begin{tabular}{>{\centering\arraybackslash} m{17em} | >{\centering\arraybackslash} m{10em}}
    \toprule
    Parameters    & Values\\ \hline
    \midrule
    Number of subcarriers, $N_\textit{c}$         & 2048         \\ \hline
    Subcarrier spacing, $\Delta f$               & 488.28 kHz     \\ \hline
    Total signal bandwidth, $B$                       & 1 GHz     \\ \hline
    CP duration,  $T_\textit{CP}$                   & 0.512 $\mu$s \\ \hline
    OFDM symbol duration w/o CP, $1/{\Delta f}$          & 2.048 $\mu$s    \\ \hline
    OFDM symbol duration w/ CP, $T_\textit{s}$ & 2.56 $\mu$s \\ \hline
    Carrier frequency, $f_\textit{c}$               & 60.98 GHz  \\ \hline
    DAC sampling rate, ${F}_{{s},{DAC}}$               & 8 GSamples/s  \\ \hline
    ADC sampling rate, ${F}_{s}$               & 1000, 500, 250, 125, and 62.5 MSamples/s  \\ \hline
    ADC bit resolution                         & 10 bits  \\
    \bottomrule
\end{tabular}
\end{table}
\begin{figure}[t!]
    \centering
    {\includegraphics[scale=0.45]{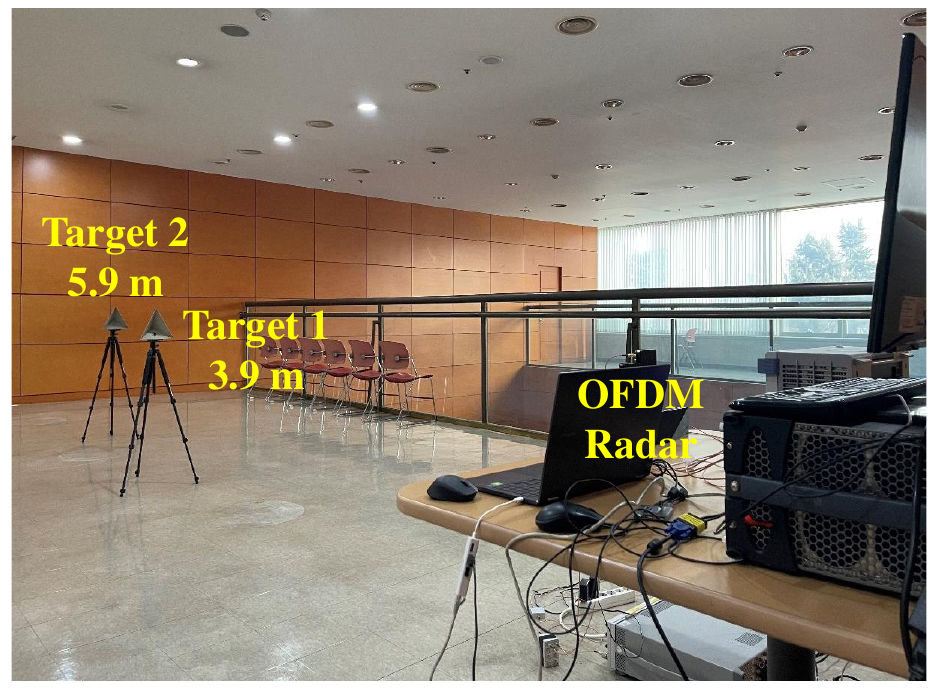}}
    \caption{Photograph of the measurement setup.}
    \label{f10}
\end{figure}
\subsection{Hardware implementation and measurement setup}
In this section, we present measurement results to verify the concept of the proposed SNS-OFDM radar scheme and the corresponding SMNC process. Similar to the simulation setup, we used an OFDM waveform of 1 GHz with 2048 subcarriers. The CP length is 0.512 $\mu$s, which equals one quarter of the length of one symbol duration. To improve the SNR, ten symbols are integrated with a coherent processing gain of 10 dB.

For the measurements, an OFDM radar system was implemented using a 60 GHz 16-channel beamforming transceiver. The OFDM signal with random data sequences is generated through a numerical tool. The OFDM signal bandwidth is 1 GHz, corresponding to a theoretical range resolution of 0.15 m with full-band processing. This is converted to analog input for the transmitter using an arbitrary wave generator (AWG). To this end, a digital-to-analog converter (DAC) in the AWG operates at 8 GSamples/s. The output of the DAC is then up-converted using the single-sideband mixer included in the transceiver module. The center frequency of the OFDM radar is set to 60.98 GHz.

The receiver of the OFDM radar system converts the RF signal to $I$ and $Q$ channels using the same frequency of the TX local oscillator (LO). After low-pass filtering, they are sampled by an oscilloscope capable of controlling various the sampling rates of the ADC. The bit resolution of the ADC is 10 bits. The sub-sampling ratios are set to $L$ = 1, 2, 4, 8, and 16, corresponding to ADC sampling rates $F_{s}$ = 1000, 500, 250, 125, and 62.5 MSamples/s. The parameters of the implemented OFDM for the measurement are summarized in Table \ref{t2}.

Two stationary targets at different ranges located 3.9 m and 5.9 m from the radar are measured. Due to the limitation of the measurement environment, there exists a wall within view of the radar at a range of approximately 9 m. A photograph of the measurement setup is shown in Fig. \ref{f10}.

\begin{figure}[t!]
    \centering
    {\includegraphics[scale=0.75]{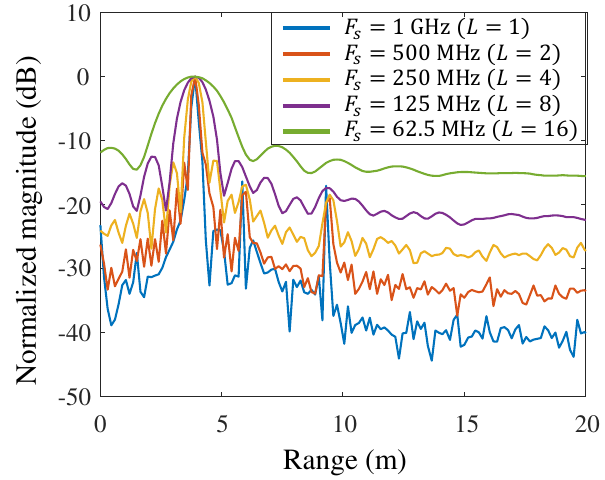}}
    \caption{Measured range profiles obtained from folded sub-band signals with various sub-sampling ratios $L$.}
    \label{f11}
\end{figure}
\begin{figure}[t!]
    \centering
    \subfloat[]{\includegraphics[scale=0.75]{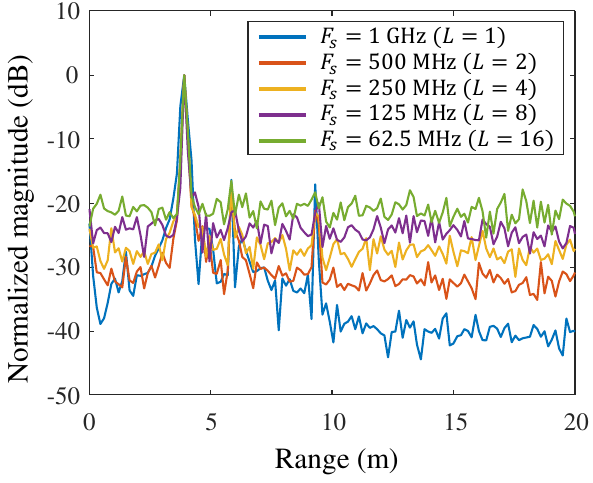}}
    
    \centering
    
    \subfloat[]{\includegraphics[scale=0.75]{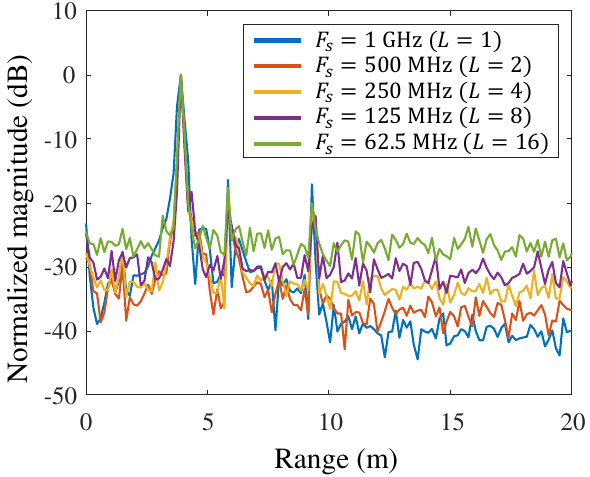}}
    \caption{Measured range profiles obtained from unfolded full-band signals with various sub-sampling ratios $L$ (a) without symbol-mismatch noise cancellation, and (b) those with symbol-mismatch noise cancellation.}
    \label{f12}
\end{figure}
\begin{figure}[t!]
    \centering
    {\includegraphics[scale=0.70]{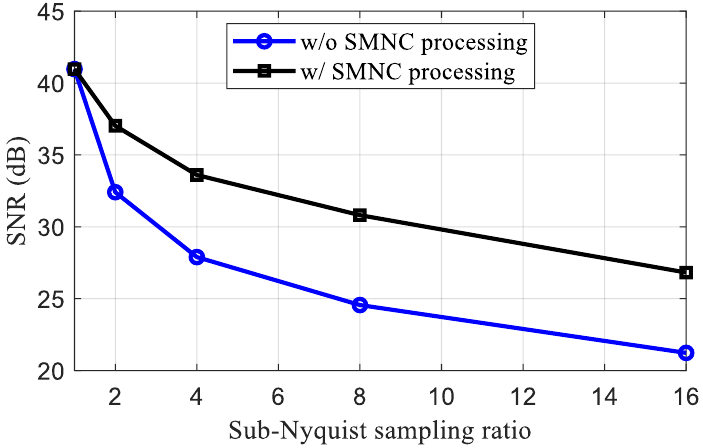}}
    \caption{Measured signal-to-noise ratio versus sub-sampling ratio $L$.}
    \label{f13}
\end{figure}
\begin{table*}[t!]
\centering
\fontsize{9}{11.5}\selectfont
\caption{Comparisons of the state-of-the-art OFDM radar systems with reduced baseband sampling rates}
\label{t3}
\renewcommand{\arraystretch}{1.5}
\begin{tabular}{>{\centering\arraybackslash} m{10em} | >{\centering\arraybackslash} m{10em}| >{\centering\arraybackslash} m{10.3em}| >{\centering\arraybackslash} m{9em}| >{\centering\arraybackslash} m{9em}}
    \toprule
    & SC-OFDM \cite{SC_OFDM1} & FC-OFDM \cite{FC_OFDM1}  & SA-OFDM$^{*}$ \cite{SA_OFDM} &  \makecell{SNS-OFDM \\ {(This work)}} \\ 
    \hline
    \midrule
ADC sampling rate  & ${B}/{L}$ & ${B}/{L}$ & ${B}/{L}$ & ${B}/{L}$    \\ \hline
DAC sampling rate & ${B}/{L}$ & ${B}/{L}$ & $B$ & $B$         \\ \hline
Max. unamb. range & $r_{max}$& ${r_{max}}/{L}$ & \makecell{${r_{max} \cdot N_a}/{N_c}$ \\ ($<{r_{max}}/{L}$)} & $r_{max}$       \\ \hline
Max. unamb. velocity & ${v_{max}}/{L}$ & $v_{max}$ & $v_{max}$ & $v_{max}$         \\ \hline
Processing gain & ${N_c N_s}$ & ${N_c N_s}/{L}$ & \makecell{$N_a N_s$ \\ $(<{N_c N_s}/{L})$} & $N_c N_s$  \\ \hline
Additional hardware & fast-settling PLL for LO stepping & frequency comb generator with $L$ carriers & $\times$ & $\times$     \\ \hline
Additional processing & phase offset calibration between sub-bands & phase offset calibration between sub-bands& $\times$ & symbol-mismatch noise cancellation \\ 
\bottomrule
\end{tabular}
\begin{minipage}{17.4cm}
\vspace{0.1cm}
\small  $^{*}$ $N_a$ $(<{N_c}/{L})$ is the number of active subcarriers determined as in earlier work \cite{SA_OFDM}.
\end{minipage}
\end{table*}
\subsection{Measurement results}
Fig. \ref{f11} shows the measured range profiles obtained from the folded sub-band signals in the SNS-OFDM radar system. Here, we utilize IFFT with $N_{c}/L$ points for range processing of the folded sub-band signal. Nyquist-sampling $L$ = 1, which covers the full band of baseband signals, clearly distinguishes the two targets and the wall, of which the ranges are 3.9 m, 5.85 m, and 9.3 m, respectively. On the other hand, the folded sub-band signal due to sub-Nyquist sampling only provides a range resolution equivalent to the sub-band bandwidth. For $L = 16$, the two targets separated by as much as 1.95 m are not resolved because the range resolution with a  bandwidth of 62.5 MHz is 2.4 m. Therefore, reducing the ADC sampling rate without proper signal demodulation in SNS-OFDM radar cannot allow one to take advantage of the full range resolution corresponding to the OFDM signal bandwidth. Furthermore, the folded sub-band involves only $N_{c}/L$ number of subcarriers, resulting in a range processing gain of $10\log_{10} ({N_{c}/L})$. Given that the folded noise also increases proportionally with the sub-sampling ratio, the detected SNR is reduced by $20\log_{10} {L}$ in the folded sub-band processing case.

Using the same measured data, we extend the folded signal to the full-band signal using the signal-unfolding process. Fig. \ref{f12}(a) shows the measured range profiles with the unfolded signals in the SNS-OFDM radar system. Compared to those obtained from the folded sub-band, the same range resolution is provided here, corresponding to 1 GHz for all sub-sampling ratios. Although full-band processing improves the range resolution, it also shows some symbol-mismatch noise compared to the results of Nyquist sampling. Due to the SMN, both the second target and the wall disappear when we adopt sub-sampling ratios of $L = 8$ and $L = 16$. This occurs because the signal magnitudes of those targets are smaller than that of the ISL raised by the SMN of the first strong target. Also, the detected target SNR in SNS-OFDM radar is much more degraded than that in the Nyquist sampling case. 

In Fig. \ref{f12}(b), the measured range profiles with the proposed SMNC processing are given, in which the SMNs are eliminated within two iterations of SMNC processing. Two targets and the wall can be estimated in the results of all sub-sampling ratios. Even in the case of $F_{s}$ = 62.5 MHz, the detected target SNRs of three peaks are 26.8 dB, 9.3 dB, and 6.8 dB. From the measurement result, it is verified that the SNS-OFDM radar system with SMNC processing is only affected by noise folding, which imposes the penalty of a 3 dB sidelobe by halving the ADC sampling rate. To demonstrate the proposed system more explicitly, the observed target SNRs versus sub-sampling ratios are computed from the measured range profiles before and after SMNC processing, as shown in Fig. \ref{f13}. The SNR improvement owing to SMNC processing is nearly identical at the sub-sampling ratios tested because the SNR before SMNC processing decreases to $10\log_{10} ({L-1})$ from (\ref{E15}) and that after SMNC processing decreases to $10\log_{10} {L}$ from (\ref{E21}). Finally, the measurement results validate the the proposed SNS-OFDM radar system both in theory and practice.            
\subsection{Discussion}
To emphasize the advantages of the proposed SNS-OFDM radar explicitly, we discuss and compare state-of-the-art OFDM radar systems with a reduced baseband sampling rate. Here, CS-based approaches are not included because they involve high computational complexity and are thus not feasible with low-cost radar sensors. Also, other radar performance metrics should be considered for a fair comparison.

First, the proposed SNS-OFDM radar can take advantage of the maximum range and velocity of a given OFDM radar system without ambiguities in the range or velocity domain. Compared to SNS-OFDM radar, the time-interleaving scheme in SC-OFDM limits the maximum detectable velocity and the frequency-interleaving scheme in FC-OFDM and SA-OFDM does so for the maximum detectable range. 

Moreover, the proposed radar system can make use of the high processing gain in one radar frame, which generates one range-Doppler map, compared to that of FC-OFDM and SA-OFDM, as the proposed system can activate all subcarriers in all OFDM symbols as opposed to the reduced number of subcarriers in the frequency-interleaving scheme. Although the processing gain of SC-OFDM is identical to that of ${N_c} {N_s}$, its frame duration is $L$ times longer than that of SNS-OFDM. 

One of the important features of SNS-OFDM is that no additional hardware is required to implement it. SC-OFDM and FC-OFDM, on the other hand, require support from LO frequency manipulation, which significantly increases the hardware complexity. Additionally, those methods are very sensitive to any phase discontinuity between the sub-bands caused by the exploitation of several LO frequencies. In contrast, with the proposed SNS-OFDM, only a symbol-mismatch noise cancellation process is necessary to remove the effect of signal folding. This approach is simple yet effective and can be executed with only a few iterations. Nevertheless, the proposed SNS-OFDM radar system as well as SA-OFDM cannot contribute to a reduction in the DAC sampling rate. Further research on reducing the DAC sampling rate in combination with SNS-OFDM will be explored elsewhere. Table \ref{t3} summarizes the key features in comparison with the other approaches.

\section{Conclusions}
In this paper, we presented a sub-Nyquist sampling OFDM radar system that effectively reduces the ADC sampling rate without requiring any additional hardware or waveform manipulations. Signal demodulation of SNS-OFDM radar that allows the unfolding of the folded signals to the full-band signal using known modulation symbols of the transmitted signal was described. We precisely analyzed the SNR of detected targets from the full-band signal. To mitigate the effects of symbol-mismatch noise and to enhance the target detection capabilities, we proposed a symbol-mismatch noise cancellation technique. This improves the detected target SNR and allows the estimation of weak targets that are not revealed without SMNC processing. Simulations and measurements with various sub-sampling ratios confirmed the effectiveness of the proposed sub-Nyquist sampling OFDM radar system, demonstrating a reduction in the ADC sampling rate according to the ratio of $L$, with a limited $10\log_{10}{L}$ increase in the noise due to noise folding. This trade-off offers significant advantages in terms of reduced hardware complexity and lower power consumption in OFDM radar. Consequently, the proposed SNS-OFDM radar system opens up promising opportunities for more power-efficient and cost-effective high-resolution digital radar systems that are applicable to various real-world applications, including autonomous vehicles, mobile robots, and intelligent buildings.

\ifCLASSOPTIONcaptionsoff
  \newpage
\fi



\bibliographystyle{IEEEtran}
\bibliography{IEEEabrv,reference}
%



%





\vfill


\end{document}